\documentclass[twocolumn,nofootinbib,superscriptaddress]{revtex4-1}
\usepackage{graphicx,epsfig}
\usepackage{mathtools}
\usepackage{amsmath}
\usepackage{epstopdf}
\usepackage{amsfonts}
\usepackage{braket}
\usepackage{fancyhdr}
\usepackage{cleveref}
\usepackage[utf8]{inputenc}
\usepackage{color}
\newcommand{\be}{\begin{eqnarray}}
\newcommand{\ee}{\end{eqnarray}}

\newcommand{\R}{{\cal R}}

\newcommand{\cd}{{\mathcal D}}
\usepackage[caption=false]{subfig}

\begin{document}

\title{Retrieving black hole information from the main Lorentzian saddle point}
\author{Cristiano Germani}
\email{germani@icc.ub.edu}
\affiliation{Institut de Ci\`encies del Cosmos, Universitat de Barcelona, Mart\'i i Franqu\`es 1, 08028 Barcelona, Spain}
\affiliation{Departament de F\'isica Qu\`antica i Astrof\'isica, Facultat de F\'isica, Universitat de Barcelona, Mart\'i i Franqu\`es 1, 08028 Barcelona, Spain}

\begin{abstract}
One of the most striking evidences of the information loss paradox is that, according to the Hawking's calculation, the correlation functions of a test scalar field exponentially decay in time. In this paper, I argue that a judicious use of the steepest descent expansion on the classical saddle point (the Black Hole background), is enough to change this early time decay into a late time growing, in agreement with information retrieval. I will explicitly show this in the Jackiw-Teitelboim gravity. There, the so-called ''ramp'' in the bulk tow-point function, is analytically obtained without the need of any other subdominant configurations of the gravity path integral.    
\end{abstract}

\maketitle

\section{Introduction}
Black Holes evaporate \cite{hawking}. According to the standard Hawking calculation they do by releasing a thermal black body spectrum which is only parameterized by the black hole's mass\footnote{For simplicity we will only discuss uncharged and no rotating black holes.} thus carrying negligible information about the quantum state of the black hole. Trusting the Hawking result until full evaporation would then lead to an information paradox \cite{paradox}, namely that quantum systems do not conserve information. 

One might think that all this information is stored inside the event horizon and for this reason inaccessible. However, Page \cite{page} has shown that trace of information should start to emerge already when the black hole is roughly evaporated half of its mass (Page time). The intriguing fact is that, at the Page time, a large black hole is still large and Hawking calculation, based on semi-classical gravity, should be correct. 

It is then clear that the Hawking calculation must contain an approximation that breaks down well before the black hole's curvature becomes large, i.e., that full quantum gravity effects become dominant.

The situation is perhaps clearer by looking at correlation functions on a Black Hole spacetime. Let us consider a scalar field. The two point function is formally defined with the following path integral\footnote{We use $\hbar=1$ units.}
\be\nonumber
\!\!\!\!\!\!\langle\phi(t',x')\phi(t,x)\rangle=N\int \cd \phi\cd g_{\mu\nu}\phi(t',x')\phi(t,x)e^{i S[\phi,g]},
\ee    
where $S$ is the action of the scalar-gravity system and $N$ the normalization\footnote{$N=\left(\int \cd \phi\cd g_{\mu\nu}e^{i S[\phi,g]}\right)^{-1}\ .$}. In the approximation of a test scalar field and low curvatures, we can consider the saddle point of the sole gravitational action, i.e. the point in which $\delta S_{gravity}=0$. This defines the semiclassical approximation. Then, we would naively expect (this is the core of Hawking's calculation)
\be\label{saddle}
\!\!\!\!\!\!\langle\phi(t',x')\phi(t,x)\rangle\simeq \bar N\int \cd \phi\,\phi(t',x')\phi(t,x)e^{i S_{scalar}[\phi,\bar g]},
\ee    
where $S_{scalar}$ is the scalar field Lagrangian evaluated in the saddle point metric $\bar g_{\mu\nu}$ and $\bar N$ is its related normalization.

To make the point, I will consider a massless scalar field and calculate the correlation function \eqref{saddle} in the vicinity of a spherically symmetric black hole horizon with metric $ds^2=-\bar g(r) dt^2+\frac{dr^2}{\bar g(r)}+r^2d\Omega_2^2$. Here, the horizon is the sphere $r=r_0$ in which $\bar g(r_0)=0$.
Expanding around that we get the Rindler metric
$ds^2\simeq - \left(\frac{2\pi}{\beta}\right)^2 x^2 dt^2+dx^2+d\ell_2^2$, 
 where $x^2=\frac{4}{\bar g'(r_0)}(r-r_0)$ and $\beta\equiv \frac{4\pi}{\bar g'(r_0)}$ is the inverse Bekenstein-Hawking temperature. Finally $d\ell_2^2=r_0^2 d\Omega^2$. To simplify the calculation I follow \cite{emparan2p}. We can consider the optical metric $\tilde g_{\mu\nu}=\frac{g_{\mu\nu}}{g_{tt}}$, the rescaled scalar $\tilde\phi=x^{-2}\phi$ and make use of the Euclidean time $t\rightarrow i \frac{\beta}{2\pi}\tau$. In this setup, we can Fourier-Bessel expand $\tilde\phi$ as
$ \tilde \phi(x,\tau)=x\sum_{n,\omega}c_{n,\omega}\psi_{\omega}(x)e^{-i 2\pi n\tau}$,
 where
 \be\nonumber
 \psi_{\omega}(\vec{x})\equiv \frac{\sqrt{2\omega \sinh\pi\omega}}{(2\pi)^{3/2}\pi}e^{-i p_i\ell^i} K_{i\omega}(p x)
 \ee 
 and $K$ is the Bessel function of second kind.
 
 Thus, around the horizon, $i S\rightarrow - S_E$ with $S_E=\sum_{n,\omega}S_E(n,\omega)$
 where
 \be \nonumber
 S_E(n,\omega)=4\pi^2  (n^2+\omega^2)c_{n,\omega}^2\ .
 \ee
 The path integral at coincident spacial points is now  Gaussian and it can be easily evaluated \cite{barbon-emparan}.
At large Lorentzian times ($t'-t\gg \beta$) \cite{barbon-rab}, one gets
 \be\nonumber
 \langle\phi(t',x)\phi{(t,x)}\rangle\xrightarrow[ (t'-t)\rightarrow \infty]{}\sim e^{-\frac{2\pi}{\beta}(t'-t)}\ ,
 \ee
 which means the correlation function decays exponentially to zero. The same can be checked for higher correlators. This clearly shows that, after long time, correlations functions as calculated above do not longer carry information. What went wrong? 
 
 In \cite{malda3d}, Maldacena proposed that after long time, other saddle points of the Euclideanized gravitational action should dominate. However, none has been found that would recover the right information so far. Lately instead \cite{island}, it has been conjectured that other configurations of the Euclidean gravitational action maximizing the entanglement entropy (the so-called islands) should dominate the very long time behavior of correlators. While this seems to work in two-dimensions \cite{island}, doubts are that it would in higher \cite{karch}. Here, I will propose a more conservative way out to the exponential decay. Before, it is useful to remind the reader some basis of the steepest descent method.
 \section{Steepest descent} \label{sd} 
 Suppose we would like to evaluate the following integral
 \be\nonumber
 I(g)=\int_{x_1}^{x_2} dx f(x) e^{-\frac{1}{g^2} h(x)}
 \ee  
 in the limit of small $g$ and in the case in which $h(x)$ has a minimum in a point $x=x_0$ (the saddle point). By shifting $x=x_0+g y$, we can expand the exponential  as
 \be
 &{}&e^{-\frac{h(x)}{g^2}}=e^{-\frac{h(x_0)}{g^2}} e^{-\frac{y^2}{2} h''(x_0)}\times\cr\nonumber
 &{}&\times(1-g\frac{y^3}{6}h'''(x_0)-y^4 h''''(x_0)-g^2\frac{3}{72}y^6h'''(x_0))^2+\ldots)\ .
 \ee
 A similar expansion can be done for the function $f(x)$
 \be\label{f}
 f(x)=f(x_0)(1+g\, y\frac{f'(x_0)}{f(x_0)}+g^2\, y^2\frac{f''(x_0)}{2 f(x_0)}+\ldots)\ .
 \ee
 In \eqref{f}, we already see something that will be crucial for the following. In the limit $f(x_0)\rightarrow 0$, the first dominant term in the steepest descent approximation is not $f(x_0)$ but rather its derivatives which, we assume not to vanish. Note that however, in this limit, the saddle point approximation is still valid. 
  
 Combining the above expansions, we get, up to ${\cal O} (g^2)$,\footnote{Odd polynomials of $y$ integrate to $0$.}
 \be\label{s1}
 \!\!\!\!\!I(g)\simeq g f(x_0) e^{-\frac{h(x_0)}{g^2}}\int_{-\infty}^{\infty}dy\, e^{-y^2\frac{h''(x_0)}{2}}(1+g^2 P(y)),
 \ee
 where $P$ is a polynomial in $y$. 
 
 Let us think about \eqref{s1} as a path integral in which $h$ plays the role of the gravitational action, $x$ of the metric, $g$ the Newtonian constant and finally $f$ the matter part. The semiclassical approximation is immediately recognized. Whenever $f(x_0)$ is not small, we can neglect the $g^2$ term and $I(g)\propto f(x_0)$. While, whenever $f(x_0)$ is exponentially small, the dominant part of the integral is
 \be\label{s2}
 I(g)\propto g^2\left(\frac{f''(x_0)}{4 h''(x_0)}-\frac{f'(x_0)h'''(x_0)}{8 h''(x_0)^2}\right)\ .
 \ee
 I'd like to pause here by stressing once more that the result \eqref{s2} has been found by using the saddle point of $h(x)$. Note however that, whenever $f\rightarrow 0$, the whole function $f(x)e^{-h/g}$ does not have a maximum in $x=x_0$ but rather, for sufficiently steep exponential, two new maximums surrounding a valley in $x=x_0$ \footnote{I thank Roberto Emparan for pointing this out.}. One might be tempted to solve the integral by considering these new peaks as new saddle points. This would be indeed correct if the curve in between the two maximums is exponentially steep. 
 
 My conjecture, that I shall prove in the two dimensional Jackiw-Teitelboim gravity (JT) \cite{JT}, is that, at least in the perturbative regime of the evaporation: 
 
 {\it The black hole information encoded in the correlations functions starts to emerge when the leading order in the steepest descent expansion of the gravitational path integral becomes of order of the next to leading order. Moreover, this information can be largely retrieved by the sole use of the main Lorentzian saddle point (the black hole background).} 
 
  \section{JT gravity}
 Two-dimensional gravity has been always the playground to capture certain properties of the significantly more difficult higher dimensional gravity. For example, one might define a black hole generated by a conical singularity where the interior is characterized by a time-like Liouville theory while the exterior by a space-like one \cite{germani}. There, the semiclassical correlators in the interior are dominated by two-equally important saddle points, this can be seen either by direct calculation or via holography \cite{germani,pereniguez}.
 
 In this paper, I will consider instead a different theory, the JT gravity\footnote{Interestingly, JT gravity and the time-like Liouville theory behave similarly is certain corner of their parameters space \cite{2006.07072}.}. The reason is that, in this theory, it has been lately argued that island configurations should be the key ingredients to recover the information in correlations functions, although, in this case one has to introduce an external topological term to the path integral ``weighting'' the different topologies (islands).
 
 Without this extra term, the JT gravity has the following action (for a full review see \cite{polchinski})
 \be\nonumber
 S_{JT}=\frac{1}{16\pi G}\left[\int d^2 x\, \Phi\sqrt{-g}\left(R+2\right)+2\int_{bdy}\Phi_b K\right]\ ,\ee 
 where $\Phi$ is the dilaton and ``bdy'' stands for boundary. The last integral is the Gibbons-Hawking term multiplied by the dilaton boundary value $\Phi=\Phi_b$. Units are such that the $AdS_2$ length is dimensionless.
 
The dilaton acts like a Lagrange multiplier fixing (at the full quantum level) the metric to be locally $AdS_2$. The connection to the black hole is done by taking the Rindler patch of $AdS_2$ and fix appropriate boundary conditions that make the choice of coordinates physical\footnote{Adding a boundary breaks the gauge invariance.}. 
Explicitly, the metric is $ds^2=-4(\rho^2-\frac{\pi^2}{\beta^2})d\tilde t^2+\frac{d\rho^2}{\rho^2-\frac{\pi^2}{\beta^2}}$ with $\Phi^2=1+\rho$. The horizons of the two copies of the black hole, are in $\rho=\pm \frac{\pi}{\beta}$. The same metric might also be written in a conformal form: After rotating to euclidean time $t_E=i \frac{2\pi}{\beta}t$ one has \cite{strominger}
  \be\label{metric}
  ds_E^2=\frac{4\pi^2}{\beta^2}\frac{dt_E^2+dz^2}{\sinh^2 \frac{2\pi}{\beta}z}\ .
  \ee
Note that the dilaton diverges at spacial infinity. We will then fix a boundary at finite distance from the horizon at $\rho=\rho_b$. 

The scalar euclidean correlators calculated in the Poincarr\'e patch of $AdS_2$, i.e.  $ds^2=Z^{-2}\left(dT_E^2+dZ^2\right)$, correspond to the Hartle-Hawking vacuum of the black hole \eqref{metric} whenever the periodicity of $T_E$ is $2\pi/\beta$ \cite{strominger}. Thus, we define, in Poincarr\'e coordinates, the boundary as the closed curve $(f(\tau),\zeta(\tau))$ parameterized by the euclidean boundary time $\tau$ \cite{malda_sch}. This curve has a fixed proper length square ($\epsilon^{-2}$) defined by the equation $f'(\tau)^2+\zeta'(\tau)^2=\frac{\zeta(\tau)^2}{\epsilon^2}$ \footnote{${}'\equiv d/d\tau$.}. The boundary value for the dilaton is $\Phi_b=\frac{\Phi_r(\rho_b)}{\epsilon}$ where $\Phi_r$ (finite for $\epsilon\rightarrow 0$) is the ``renormalized'' dilaton. For $\epsilon\rightarrow 0$ we recover full $AdS_2$. In the following, we will only consider the leading order in small $\epsilon$, thus $\zeta(\tau)\simeq\epsilon f'(\tau)$. 
  
At this order we get \cite{malda_sch}
  \be\label{sch}
  S_{JT}\xrightarrow [\epsilon\rightarrow 0]{}\frac{i}{2 g^2} \int d\tau\, {\rm Sch}(f, \tau)\ ,
  \ee
where
\be{\rm Sch}(f,u)=-\frac{1}{2}\frac{f''^2}{f'^2}+\left(\frac{f''}{f'}\right)'
\ee 
is the Schwarzian derivative of $f=f(\tau)$. In \eqref{sch} I have re-defined $g^2\equiv \frac{4\pi G}{\bar\Phi_r}\ll 1$. 
  
Varying this action with respect to $f$ \footnote{We only consider disk topologies.} we find the periodic solution $f=\tan(\frac{\tau}{2})$ which corresponds to the black hole configuration. Indeed, the bulk metric might be equivalently written as \cite{rodes} $ds^2=\frac{\dot f(u)\dot f(v)}{(f(u)-f(v))^2} du dv$, where $u=i\frac{2\pi}{\beta}(t+z)$ and $v=i\frac{2\pi}{\beta}( t-z)$ are now bulk ''euclidean'' coordinates. With the euclidean saddle solution $f=\tan\frac{\tau}{2}$, we recover the black hole metric \eqref{metric}. It is then clear that the gravitational degree of freedom is completely mapped into a function $f$ of the boundary time. 
 
\subsection{Adding matter}
In this section, for simplicity but without loss of generality, I will consider a massless scalar field and calculate its two-point function. Alternatively, as it is done repeatedly in the literature, one may calculate its dual correlator living at the boundary of the theory (from now on dual correlator). The full numerical calculation of this exists in the related one-dimensional Sachdev-Ye-Kitaev theory (SYK) \cite{SYK} (see e.g. \cite{matrix}). There, one finds an exponential decay of the dual correlation function superseded by a linear growing (the ''ramp''), in compatibility with the Page curve. However, the literature interpretation of this result in the path integral language (see e.g. \cite{saad, rodes}) differs greatly from what I shall present here. In \cite{saad, rodes}, by using analytical methods (see also \cite{others} for alternative techniques) the information recovery is attributed to new non-saddle gravitational configurations of the Euclidean action weighted by an external topological term related to the Euler characteristic of the configuration chosen. Here instead, I claim that the information retrieving behavior of the bulk correlators is just the result of a correct implementation of the steepest descent method around the Lorentzian saddle point: the classical black hole geometry.  
  
  The rest of the paper is about calculating the following two-point correlation function
  \be\nonumber
  \langle\phi(t',z)\phi(t,z)\rangle= N\int \cd \phi\cd f\, \phi(t',z)\phi(t,z) e^{i S_{\phi}(\phi,f)}e^{i S_{JT}}
  \ee
  where $S_{\phi}$ is the action of the massless scalar $\phi$ and $N$ the usual normalization. I will follow the procedure outlined in \cite{rodes}. Firstly, we will do the path integral of $\phi$ for a given configuration of $f$. By noticing that different $f$ are related to different time reparameterizations of $AdS_2$, one can use the well known scalar correlators in $AdS_2$ and insert them into the path integral for $f$. One finds (see \cite{rodes} for details)
  \begin{widetext}
  \be\label{path}
  \langle\phi(t',z)\phi(t,z)\rangle=-\int_v^u d\tau_1\int_{v'}^{u'}d\tau_2\, N \int\cd f \frac{\dot f(\tau_1)\dot f(\tau_2)}{(f(\tau_1)-f(\tau_2))^2}e^{-\frac{1}{2g^2}\int du\, {\rm Sch}(f,u)}\ .
  \ee  
\end{widetext}
  The measure $\cd f$ of this path integral requires some discussion \cite{witten}. First of all, the path integral has redundant configurations due to the invariance of the action with respect to the $SL(2,\R)$ group. Thus, we need to factor out those configuration to get a finite result. Defining $f=\tan(\psi(u)/2)$, the natural measure on the disk is the Pfaffian one $\cd f\rightarrow\prod_u\frac{d\psi(u)}{\psi'(u)}$. By using a new (Majorana) fermionic variable $\eta=d\psi/\psi'$, one finds that the path integral \eqref{path} may be recast into \cite{witten}
  \begin{widetext}
  \be\label{2pt}
 \langle\phi(t',z)\phi(t,z)\rangle=-\int_v^u d\tau_1\int_{v'}^{u'}d\tau_2 N \int\frac{\cd \psi \cd\eta}{SL(2,\R)} \frac{\dot f(\tau_1)\dot f(\tau_2)}{(f(\tau_1)-f(\tau_2))^2}e^{-\frac{1}{2}\int du\, \left(\frac{\psi''^2}{g^2 \psi'^2}-\frac{\psi'^2}{g^2}+\frac{\eta''\eta'}{\psi'^2}-\eta'\eta\right)}\ ,  
  \ee 
  \end{widetext}
 where $\cd \psi$ and $\cd \eta$ are now flat measures \cite{flat}. Finally, I define $G(\Delta t,z)\equiv{\rm Re}  \left(\langle\phi(t',z)\phi(t,z)\rangle\right)$ with $\Delta t=t'-t$.
 
 \section{Recovering the ``ramp'' in the steepest descent approximation}  
 I will now show that the equivalent to the ramp found numerically in \cite{matrix} for the boundary correlator, precisely comes from the order $g^2$ in the steepest descent approximation of the path integral in the bulk correlator.  
 
 We can expand $\psi=\tau+g\gamma(\tau)$ in \eqref{2pt} and follow the steps of Sec. \ref{sd}. At zeroth order in $g$ it is very easy to get
 \be\nonumber
 \langle\phi(t',z)\phi(t,z)\rangle\xrightarrow[\frac{2\pi}{\beta}\Delta t\rightarrow \infty]{}\left(1-\cosh\frac{4\pi z}{\beta}\right)e^{-\frac{2\pi}{\beta}\Delta t},
 \ee 
 where $\Delta t\equiv (t'-t)>0$.
 
 The ${\cal O}(g^2)$ is instead more complicated. Expanding \eqref{2pt} we get
 \begin{widetext}
 \be
 G(\Delta t,z)&=&\langle\phi(t',z)\phi(t,z)\rangle\Big|_{g=0}-\frac{g^2}{8}\int_u^vd\tau_1\int_{v'}^{u'} d\tau_2\csc\left(\frac{\tau_1-\tau_2}{2}\right)^4\times\Big(\langle\gamma'(\tau_1)\gamma'(\tau_2)\rangle\left(1-\cos(\tau_1-\tau_2)\right))+\cr&+&\left(\langle\gamma(\tau_1)^2\rangle+\langle\gamma(\tau_2)^2\rangle\right)\left(1-\frac{1}{2}\cos(\tau_1-\tau_2)\right)-\langle\gamma(\tau_1)\gamma(\tau_2)\rangle\left(\cos(\tau_1-\tau_2)+2\right)-\cr
 &-&\sin(\tau_2-\tau_1)\left(\langle\gamma(\tau_1)\gamma'(\tau_1)\rangle+\langle\gamma(\tau_2)\gamma'(\tau_2)\rangle+\langle\gamma(\tau_1)\gamma'(\tau_2)\rangle-\langle\gamma(\tau_2)\gamma'(\tau_1)\rangle\right)+{\cal O}(\gamma^4)\Big)\ .
 \ee
 \end{widetext}
The correlators of $\gamma$ can be calculated by expanding $\gamma=\sum_{n}e^{-in\tau}k_n$ and imposing the reality conditions. They will result on Gaussian integrals. See for example \cite{chinese}. Similarly for the fermionic variable $\psi$ (see e.g. \cite{witten}).
 
\begin{center}  
\begin{figure}[h]
	\centering
	\begin{tabular}{cc}
		\includegraphics[scale=0.8]{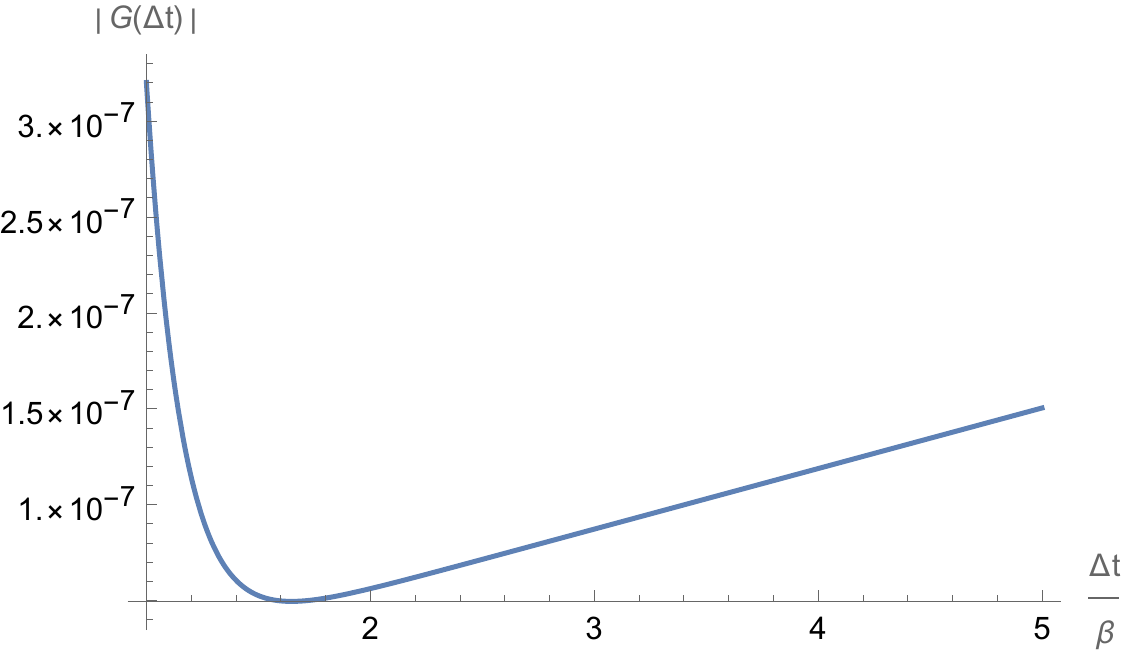} &	\includegraphics[scale=0.8]{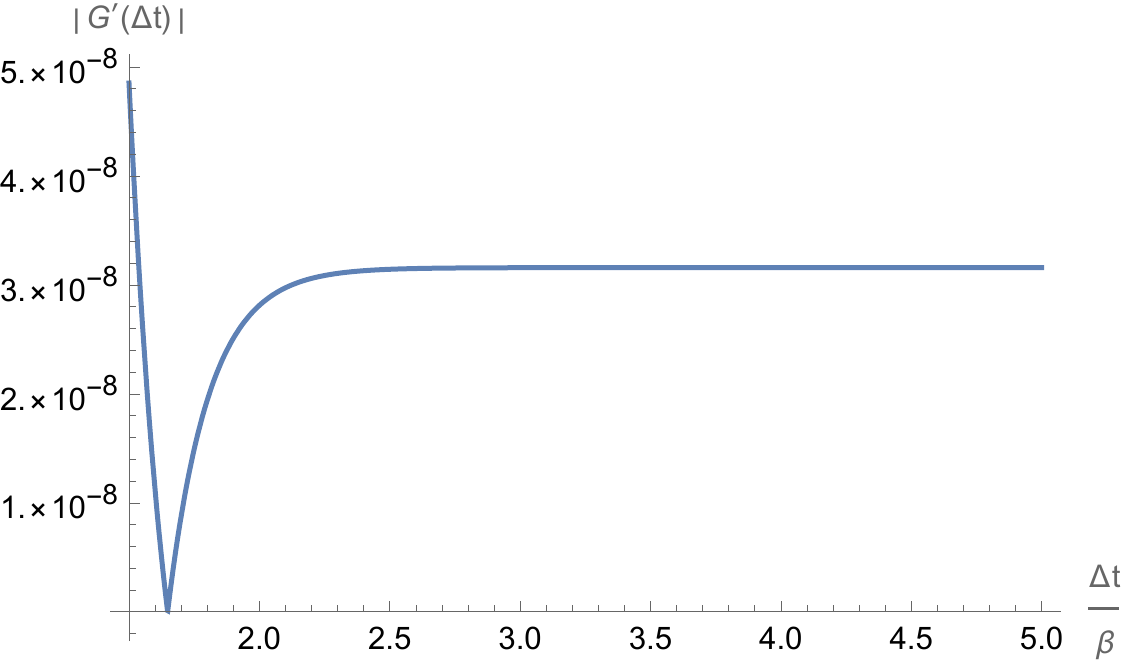} \\
	\includegraphics[scale=0.8]{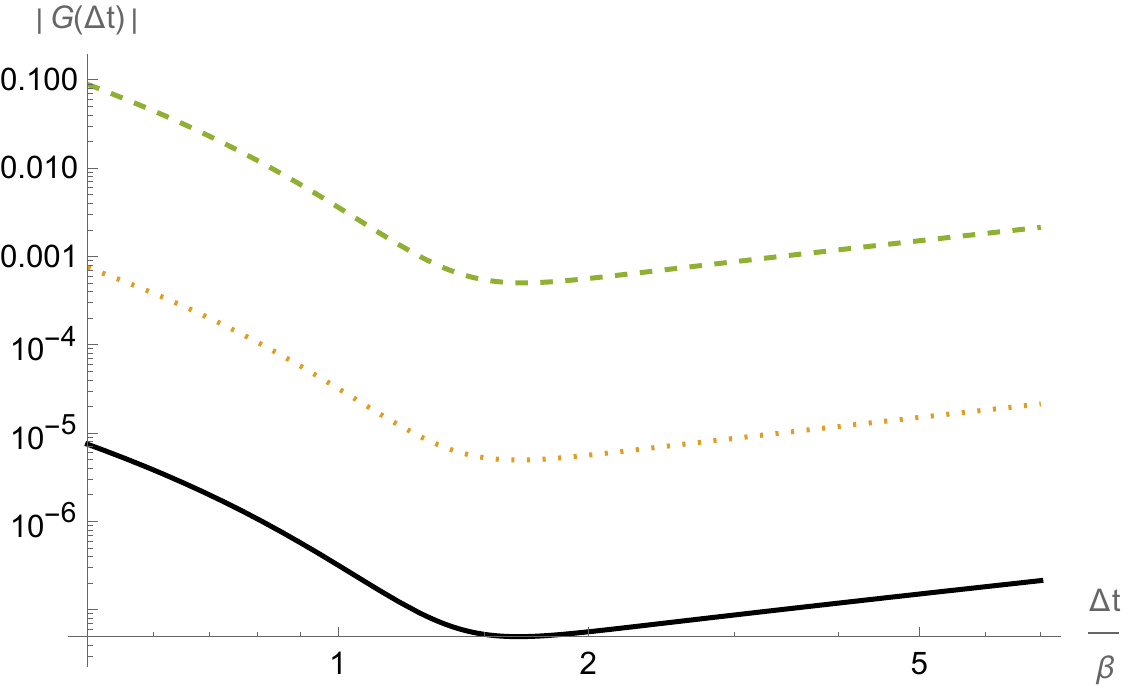} &
	\includegraphics[scale=0.8]{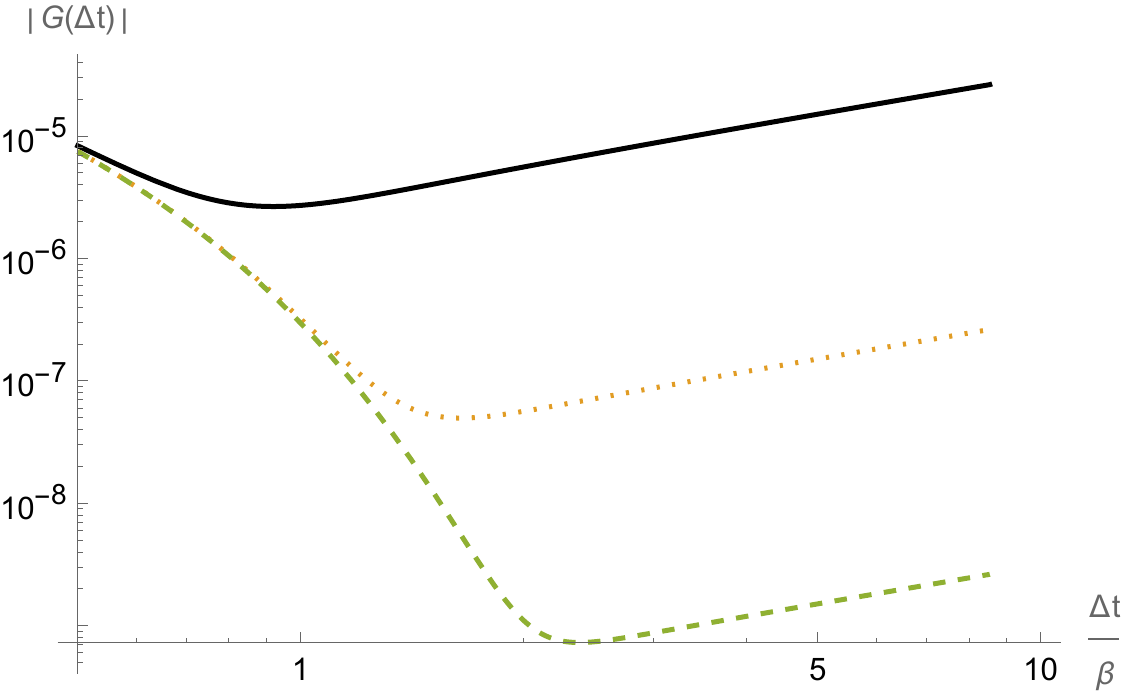}
	\end{tabular}
\caption{First line: correlation function and its time derivative for $z=0.1$ and $g=0.01$. We clearly see the emergence of the ramp after the exponential decay. Second line, first plot: correlations functions for different values of $z$, namely: (solid) $z=0.1$, (dotted) $z=0.01$, (dashed) $z=0.001$ and $g=0.01$. Second plot second line, $z=0.1$ and: (solid) $g=0.1$, (dotted) $g=0.01$, (dashed) $g=0.001$. We note that the dependence on $z$ of the dip time is very mild.}\label{fig1}
\end{figure}
\end{center}

Interestingly enough the higher correlators (which are still at order $g^2$) sum to zero. The analytical form of $G$ is rather long and not very illuminating. In Fig. \ref{fig1}, I have plotted the correlation function $G$ for different values of $g$ and $z$. There, it is clear that the two-point correlation function smoothly join a linear grow after an exponential decay. The asymptotic linear growing is 
\be
\Big|G(\Delta t,z)\Big|\sim 32\pi^2 g^2 z^2\Delta t\ .
\ee
It is important to stress that this correlator {\it is not} the one-loop correction to the dual operator calculated, for example, in \cite{malda_sch}. Here, I have worked out the bulk two-point function of the mass-less scalar at order $g^2$ and thus no input from the $AdS/CFT$ conjecture have been used. Moreover, while in the calculation of the dual operators there is an ambiguity on defining the Wick rotation back to Lorentzian time, here, as a result of the $u$ and $v$ integration before Wick rotation back to Lorentizian time, this ambiguity disappears.

Finally, one may ask whether the ${\cal O}(g^4)$ terms, in the perturbative $g\ll 1$ regime, might also be important whenever the ${\cal O}(1)$ term in $g$ becomes subdominant. The latter happens because of the well known exponential suppression in time due to the presence of an horizon and thus, it is not parametric in $g$. On the contrary, because the ${\cal O}(g^2)$ is polynomially growing in time, the ${\cal O}(g^4)$ is bound to be parametrically smaller than the ${\cal O}(g^2)$, at least for the initial part of the ramp. Eventually though, the parametric suppression might be overtaken by the temporal growing of the correlation functions. This is where we expect a plateau behavior \cite{saad}.

\section{Conclusions and outlook}

By considering the next to leading order expansion in the steepest descent approximation, I have shown that, at least in the two-dimensional JT gravity, the two-point correlation function in a black hole spacetime, change its exponential decay into a linear grow. This can be immediately compared, at least qualitatively, with the numerical findings in the SYK theory by taking into account that this dual theory lives at the point $z=z_b\simeq \frac{1}{2\rho_b}\ll 1$ of the JT black hole geometry. 

It is important to stress that the next to leading order ''correction'' in steepest descents introduced here, is dominant in path integral calculations (gravitational or not), whenever the correlators exponentially decay to zero at the leading order in saddle points. Thus, whether this solves the information paradox or not, it has to be implemented to higher dimensional gravitational theories in the presence of black holes, which I postpone for future work. 

Coming back to the JT gravity, I have found that at a finite time $\Delta t$, one finds that $G(\Delta t,z_b)\propto z_b^2$. Thus the dip in the boundary two-point correlation function connecting the exponential decay with the linear grow, is suppressed by the large dilaton (note that, conversely from e.g. \cite{saad}, here I work with normalized partition functions). Moreover, while this dip happens numerically at the bulk time of order $\Delta t^{dip}\sim\frac{\beta}{2\pi}\log(\frac{\pi} {g^{2}})$, as one would expect from Page arguments, from the point of view of the boundary Lorentzian time ($\tau_L$) on which SYK theory is constructed, we have $\tau^{dip}_L\sim 4\pi\frac{(4\pi G)^2}{\epsilon g^4\beta}\Delta t^{dip}\sim \frac{4\pi G}{g^2}\Phi_b\ln(\frac{\pi}{g^2})$.  

The boundary correlation function calculated in this paper entirely comes from the Lorentzian saddle point of the path integral and reproduces, at least qualitatively, the dual one of \cite{saad,rodes} calculated from the so-called ''trumpet'' configurations. This seemly paradoxical fact might have a simple explanation. As discussed in Sec. \ref{sd}, the saddle point of the gravitational action, as seen from the full path integral of matter + gravity, becomes a valley after the exponential decay of the matter part. This, in turn, might well generate two neighbor peaks that, in terms of gravity configurations, could be related to the trumpets.

Finally, we can expect that the linear growing of the correlation function should change into something else whenever the scalar back-reactions becomes important, i.e. towards the end of the black hole's life \cite{andreas}. Indeed, in the numerical calculations of \cite{saad}, the ramp eventually encounters a plateau. This plateau might still be found as a Lorentzian saddle point calculation where back-reactions are also included. I leave the proof of this statement for future work. 
\onecolumngrid
\twocolumngrid
 \begin{acknowledgments}
I would like to thank Thomas Mertens and Andreas Blommaert for useful explanations of their work and for their comments on the main results of this paper. Moreover, I thank Jaume Garriga for a discussion on soft mode propagators in the Schwarzian theory and Bartomeu Fiol for interesting comments on the first draft of this work. Finally I would like to thank Roberto Emparan for constructive criticisms on the saddle point approximation. I am partially supported by the Unidad de Excelencia Maria de Maeztu Grant No. CEX2019-000918-M, and the Spanish national grants PID2019-105614GB-C22, PID2019-106515GB-I00.
\end{acknowledgments}


\end{document}